\documentclass[reprint, superscriptaddress, amsmath,amssymb, aps]{revtex4-1}
\usepackage{graphics}
\usepackage{graphicx}
\usepackage{subfigure}
\usepackage{dcolumn}
\usepackage{bm}
\usepackage{hyperref}
\usepackage{color}

\begin{document}

\title{Topologically protected second-harmonic generation via doubly resonant high-order photonic modes}

\author{Yafeng Chen}
\thanks{These authors contribute equally to this work}
\affiliation{State Key Laboratory of Advanced Design and Manufacturing for Vehicle Body, Hunan University, Changsha, Hunan 410082, China}
\affiliation{Department of Mechanical Engineering, Hong Kong Polytechnic University, Hong Kong SAR, China}

\author{Zhihao Lan}
\thanks{These authors contribute equally to this work}
\email{z.lan@ucl.ac.uk}
\affiliation{Department of Electronic and Electrical Engineering, University College London, London WC1E 7JE, United Kingdom}

\author{Jensen Li}
\affiliation{Department of Physics, The Hong Kong University of Science and Technology, Clear Water Bay, Kowloon, Hong Kong, China}

\author{Jie Zhu}
\email{jiezhu@tongji.edu.cn}
\affiliation{Institute of Acoustics, School of Physics Science and Engineering, Tongji University, Shanghai 200092, China}

\begin{abstract}
Topology-driven nonlinear light-matter effects open up new paradigms for both topological photonics and nonlinear optics. Here, we propose to achieve high-efficiency second-harmonic generation in a second-order photonic topological insulator. Such system hosts highly localized topological corner states with large quality factors for both fundamental and second harmonic waves, which could be matched perfectly in frequency by simply tuning the structural parameters. Through the nonlinear interaction of the doubly resonant topological corner states, unprecedented frequency conversion efficiency is predicted. In addition, the robustness of the nonlinear process against defects is also demonstrated. Our work opens up new avenues toward topologically protected nonlinear frequency conversion using high-order photonic topological modes.
\end{abstract}
\maketitle

\section{\label{sec:intro}Introduction}
Photonic topological insulators (PTIs) featured with gapless topological edge states that are resilient to impurities provide promising opportunities for the advanced manipulation of light \cite{Khanikaev17NatPho_2Dtopo, WuHu15PRL, Hang18PRL_ExpSpin, Bahari17Science_lasing,yafeng RRL, Harari18Science_TLlaserT, Zeng20Nature_valleylaser, Ozawa19RMP, Lu16NP, Khanikaev13NM_PTI, Lu14NP_review}. Recently, the high-order PTI has been theoretically predicted and experimentally realized \cite{Xie18PRB_corner, Xie19PRL_exp, Chen19PRL_exp, Hassan19NatPho_coner, Chen20PRR_inverse, Kim20Light, Mittal10NatPho_quadrupole, He20NatCom_Quadrupole, Zhang20AdvSci}.  In two-dimensional (2D) systems, the second-order PTI (SPTI) hosts gapped edge states and gapless corner states, which goes beyond the traditional bulk-edge correspondence. Based on tightly localized corner states, high quality (Q) factor topological nanocavity \cite{Ota19Optica_cornercavity} and topological nanolasers \cite{Zhang20Light_cornerlaser, Kim20NatCom_conerlasing} have been experimentally realized. 

On the other hand, ongoing efforts have been directed toward connecting topological modes and nonlinear light-matter effects as their interactions could lead to new physics not possible for linear photonic topological systems \cite{Smirnova20APR_NTPreview, Dobrykh18PRL_nonlinear, Smirnova19PRL_THG, Mukherjee_Science20, Maczewsky_Science20, Xia_Science21, Jurgensen_arxiv21_Thouless}, including nonlinear control \cite{Dobrykh18PRL_nonlinear} and imaging \cite{Smirnova19PRL_THG} of photonic topological edge states, the formation of topological solitons \cite{Mukherjee_Science20}, nonlinearly-induced topological transitions \cite{Maczewsky_Science20},  nonlinear tuning of non-Hermitian topological states \cite{Xia_Science21} and nonlinear Thouless pumping \cite{Jurgensen_arxiv21_Thouless}.
 Nonlinear frequency conversion is one of the fundamental nonlinear optical processes \cite{Boyd08Book}. It has been demonstrated that the presence of topological edge states could enhance the harmonic generation based on robustness of the local field enhancement \cite{Kruk19NatNaT_nonlinear, Wang19NatCom_harmonic, Qian18OE_doubleedge, You20SciAdv, Lan21PRA_valleySHG, Lan20PRB_nonlinear}. One of the main limiting factors for high conversion efficiencies is the Q factors of the resonant modes of the devices \cite{Carletti18PRL_bic, Koshelev20Science_bic, Wang20Optica_bicSHG, Minkov19Optica_bicSHG,ZinlinSHG}. Essentially, SPTIs with tightly localized corner states could be deemed as topology-protected resonant cavities with high Q factors \cite{Ota19Optica_cornercavity}, thus providing a promising platform for improving the harmonic generation efficiency. Recently, enhanced third-harmonic generation from topological corner states has been demonstrated \cite{Kruk20NanoLett_cornerSHG}. However, the system only supports one corner state for the fundamental wave (FW), thus lacking the nonlinear optical interaction between topological corner states and limiting the conversion efficiency. For the giant enhancement of light-matter interaction, it is highly desirable to engineer two topological corner states for both the FW and the second-harmonic (SH), which is extremely challenging and has not been yet realized. 

Here, we demonstrate topologically protected high-efficiency second-harmonic generation (SHG) in an all-dielectric SPTI. The key novelty of our work lies in the design of a SPTI hosting topological corner states within two band gaps, which could be matched perfectly in frequency. We show that the nonlinear interaction of the doubly resonant topological corner states produces order-of-magnitude enhancement of the SHG efficiency. Moreover, since the corner states are topologically protected, the SHG is robust against defects, a functionality that nontopological nonlinear optical devices cannot offer.

\begin{figure}
\includegraphics[width=\columnwidth]{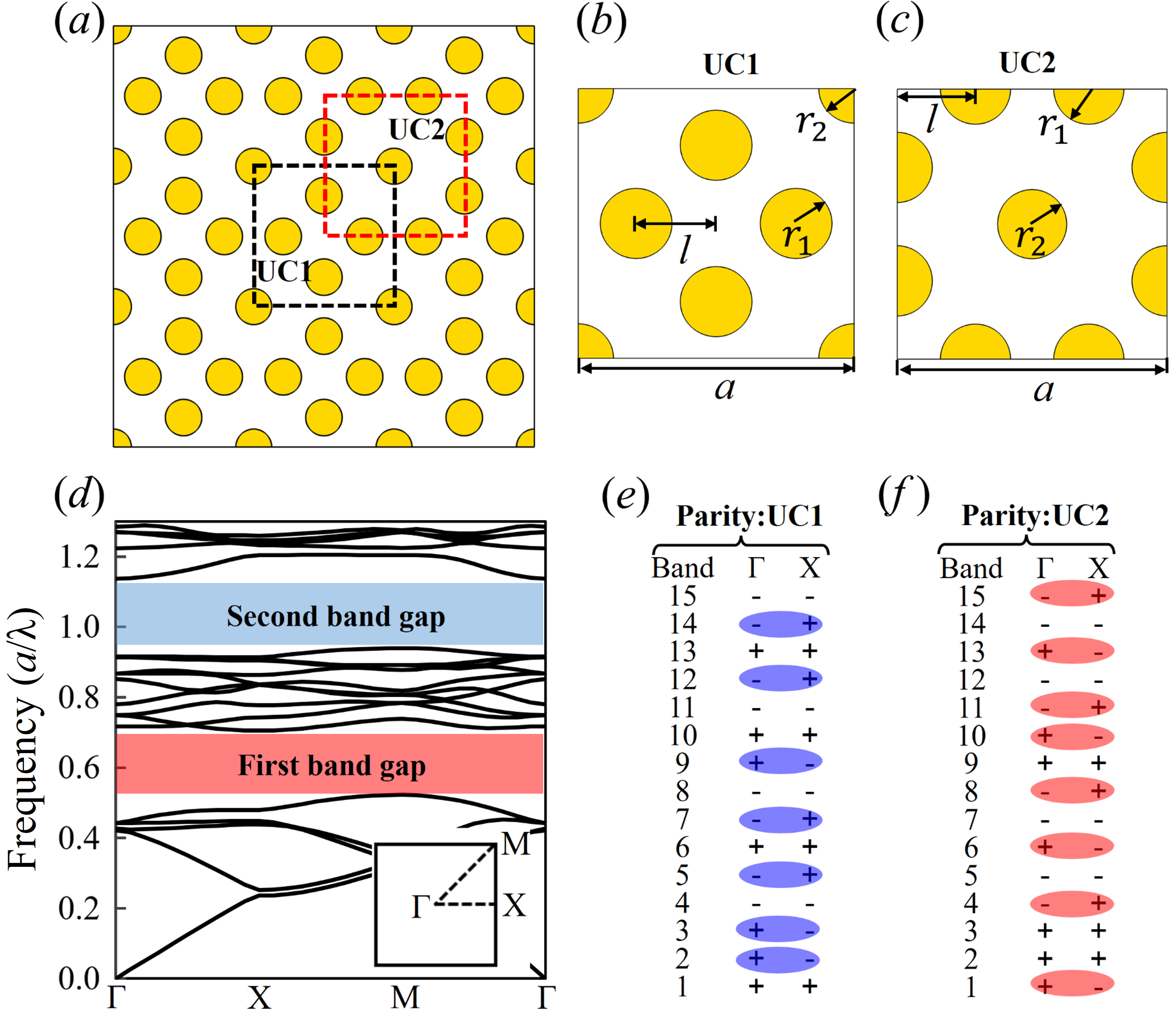} 
	\caption{\textit{Existence of two topological band gaps in a 2D dielectric PC for SHG.}  (a) Schematic of the PC and two different choices of the unit cell, denoted as UC1 and UC2. (b) UC1 and (c) UC2 contain two different kinds of cylinders with radius of $r_1$ and $r_2$, where $l$ is the distance of the $r_1$ cylinder with respect to the center of UC1 (or corner of UC2) and $a$ is the lattice constant. (d) The TM band diagram of the PC at $r_1=0.13a$, $r_2=0.128a$ and $l=0.29a$, which hosts two topological band gaps. The insert shows the first Brillouin zone of the PC with the high symmetry points marked by $\Gamma$, $X$ and $M$. (e) and (f) Parities of the 15 energy bands below the second band gap at the high symmetry points $\Gamma$ and $X$ for UC1 and UC2, respectively. 	\label{fig:fig1} }
\end{figure}

\section{\label{sec:design}Design of the system}  
We consider a 2D square photonic crystal (PC) with the $C_{4v}$ point group symmetry, see Fig. \ref{fig:fig1}(a). The PC is made of dielectric cylinders with dielectric constant of $\epsilon=12.25$ and second-order nonlinear susceptibility of $\chi^{(2)}=10^{-21}\textrm{CV}^{-2}$. We focus on the transverse magnetic (TM) modes of the PC herein and for simplicity, the frequency $f$ is normalized with respect to $a/\lambda$. Previous studies of SPTIs based on the 2D Su-Schrieffer-Heeger model suggest that choosing the unit cell (UC) in different ways can lead to distinct topological behaviors of the system \cite{Xie18PRB_corner, Xie19PRL_exp, Chen19PRL_exp, Hassan19NatPho_coner, Ota19Optica_cornercavity, Chen19PRB_cornerbic, Liu17PRL_zeroBC, Meng20APL_3D}. To create topological band gaps, we select two different UCs (UC1 and UC2) from the PC (Fig. \ref{fig:fig1}(a)), whose configurations are given in Figs. \ref{fig:fig1}(b) and \ref{fig:fig1}(c), respectively. Note that the PC contains two different kinds of cylinders with radius $r_1$ and $r_2$ and UC2 is a translation of UC1 by ($a/2, a/2$) with $a$ the lattice constant. A typical band diagram of the PC at $r_1=0.13a$, $r_2=0.128a$, and $l=0.29a$ is presented in Fig. \ref{fig:fig1}(d), which hosts two band gaps. While the first band gap locates between bands 5 and 6 and spans a frequency window from $0.523a/\lambda$ to $0.705a/\lambda$, the second band gap locates between bands 15 and 16 and spans from $0.940a/\lambda$ to $1.138a/\lambda$. Note that while the band diagram in Fig. \ref{fig:fig1}(d) is the same for both UC1 and UC2 as they encode the same PC, the symmetry properties of the bands, especially at the high symmetry points in the first Brillouin zone (see the insert of Fig. \ref{fig:fig1}(d)), e.g., $\Gamma$ and $X$, can be different for UC1 and UC2, which will result in distinct topological behaviors for UC1 and UC2.

\section{\label{sec:Topological-properties}Topological properties of the two band gaps}  
The topological properties of the two band gaps for UC1 and UC2 can be analyzed through the 2D polarization $\mathbf{P}=(P_x,P_y)$ defined by \cite{Liu19PRL_Helical, Zhang19PRL_nonHsonic, Zhang19AdvMat_Acoustic},  
\begin{gather}
P_i=\frac{1}{2} \left( \sum_n q_i^n \textrm{modulo}  2 \right), (-1)^{q_i^n}=\frac{\eta_n(X_i)}{\eta_n(\Gamma)},
\label{2Dpol}
\end{gather}
where $i=x,y$ represents the direction, $\eta_n$ is the parity of the n-th band at the high symmetry point of the first Brillouin zone (see the insert of Fig. \ref{fig:fig1}(d)) and the summation over $n$ is for all the bands below the band gap (first or second).  Note that due to the $C_{4v}$ point group symmetry of our PC structure, $P_x=P_y$ and moreover, the parities of the n-th band at the high symmetry points of $\Gamma$ and $X$ can be identified by the behavior of the corresponding eigenmode profile under the inversion operation with respect to the center of the unit cell. The corresponding eigenmode profiles used for identifying the parities are given in the Figs. 2-5, where the $p$ modes have an odd parity (-), whereas the $s$ and $d$ modes have an even parity (+). The results of the parity distribution at the high symmetry points of $\Gamma$ and $X$ for UC1 and UC2 are given in Figs. \ref{fig:fig1}(e) and \ref{fig:fig1}(f), respectively. Taking the results of Figs. \ref{fig:fig1}(e) and \ref{fig:fig1}(f) into Eq. (\ref{2Dpol}), the 2D polarization can be determined to be $P=(1/2, 1/2)$ for both the first and the second band gap of UC1, and $P=(0,0)$ for those of UC2, meaning that the two band gaps in UC1 are both topologically nontrivial while they are both trivial in UC2. Eq. (\ref{2Dpol}) also implies a simple rule to identify the topological property of the band gap through the number of pairs of parity with opposite signs at $\Gamma$ and $X$ for the same band, i.e., while an odd number of pairs indicates nontrivial topology, an even number implies a trivial gap. For example, for UC1, there are 3 pairs below the first band gap (with 5 bands) and 7 pairs below the second band gap (with 15 bands), thus the two band gaps are both topologically nontrivial for UC1.  

\begin{figure}
\includegraphics[width=\columnwidth]{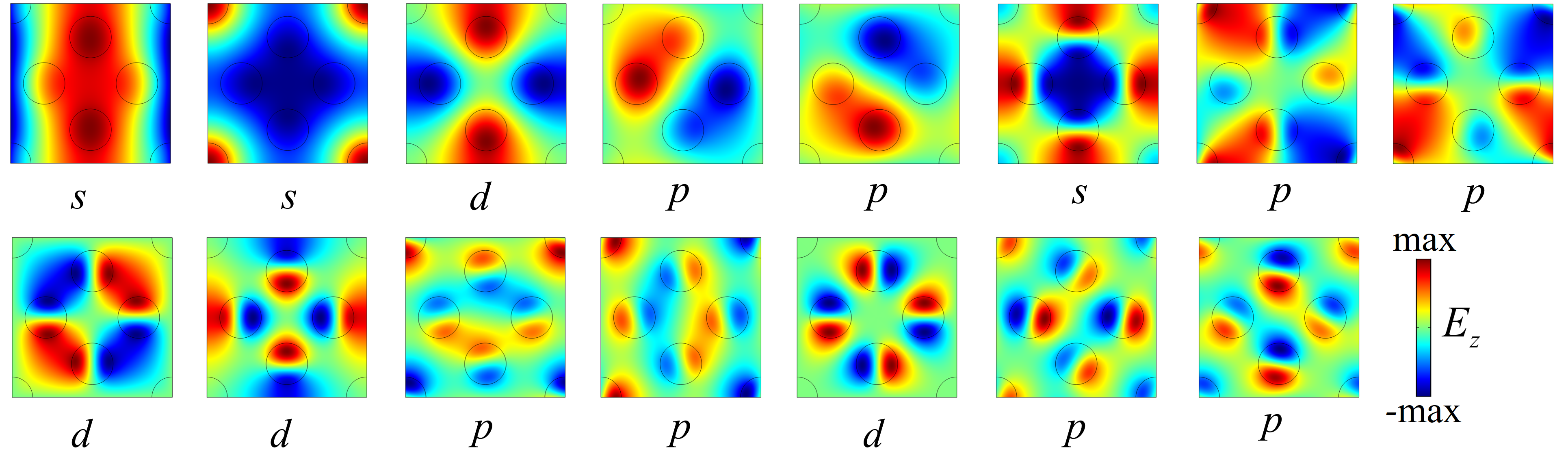} 
\caption{{Eigenmodes at high symmetry point of $\Gamma$ for UC1. Upper panel: band 1-8 (from right to left). Bottom panel: band 9-15 (from left to right).} \label{fig:fig2}}
\end{figure}

\begin{figure}
\includegraphics[width=\columnwidth]{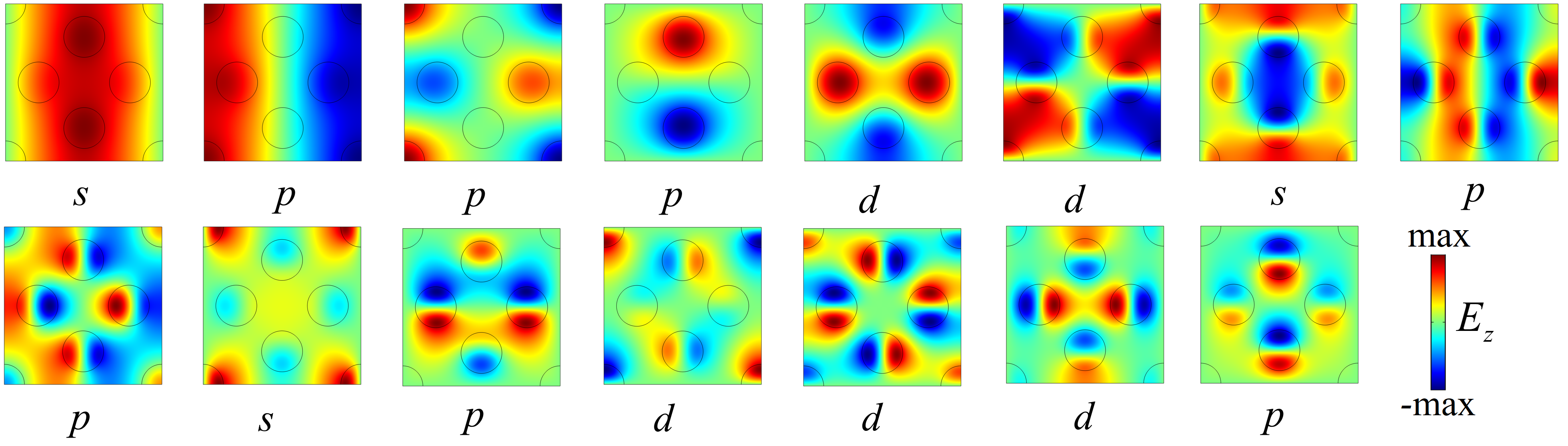} 
\caption{{Eigenmodes at high symmetry point of $X$ for UC1. Upper panel: band 1-8 (from right to left). Bottom panel: band 9-15 (from left to right)}. \label{fig:fig3}}
\end{figure}

\begin{figure}
\includegraphics[width=\columnwidth]{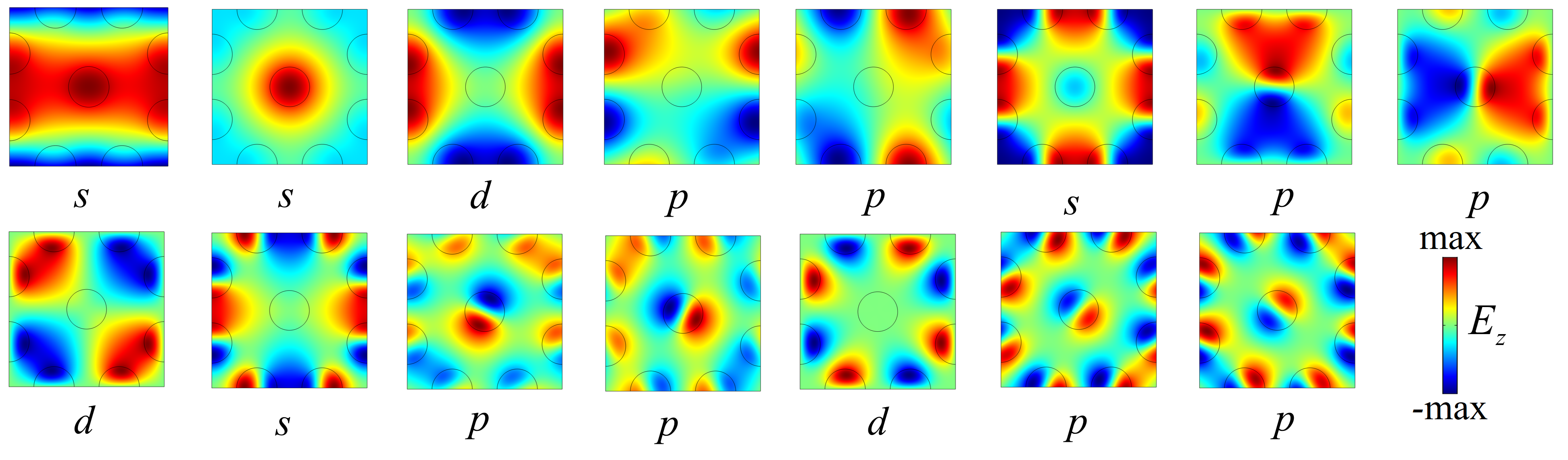} 
\caption{{Eigenmodes at high symmetry point of $\Gamma$ for UC2. Upper panel: band 1-8 (from right to left). Bottom panel: band 9-15 (from left to right).} \label{fig:fig4}}
\end{figure}

\begin{figure}
\includegraphics[width=\columnwidth]{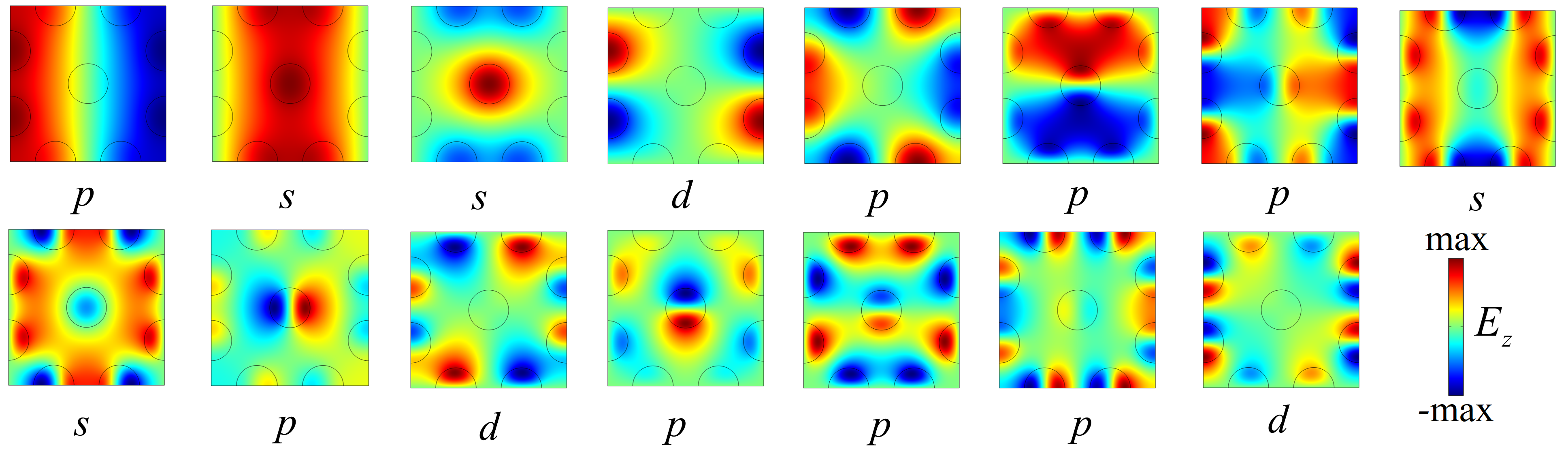} 
\caption{{Eigenmodes at high symmetry point of $X$ for UC2. Upper panel: band 1-8 (from right to left). Bottom panel: band 9-15 (from left to right).} \label{fig:fig5}}
\end{figure}

According to the bulk-edge correspondence principle, a PC with nonzero $(P_x, P_y)$ will lead to the presence of topological edge states at its interface with another PC having zero $(P_x, P_y)$ (see Supplementary Material~\cite{supple_material}). Meanwhile, the co-existence of nonzero $P_x$ and $P_y$ could induce a topological corner charge \cite{Liu19PRL_Helical},
\begin{gather}
Q=4P_xP_y.
\end{gather}
As such, the corner charges of the two band gaps are both 1 for UC1 and 0 for UC2, ensuring the existence of topological corner states within both band gaps at a $90^\circ$ corner between trivial and nontrivial regions.

\begin{figure}
\includegraphics[width=\columnwidth]{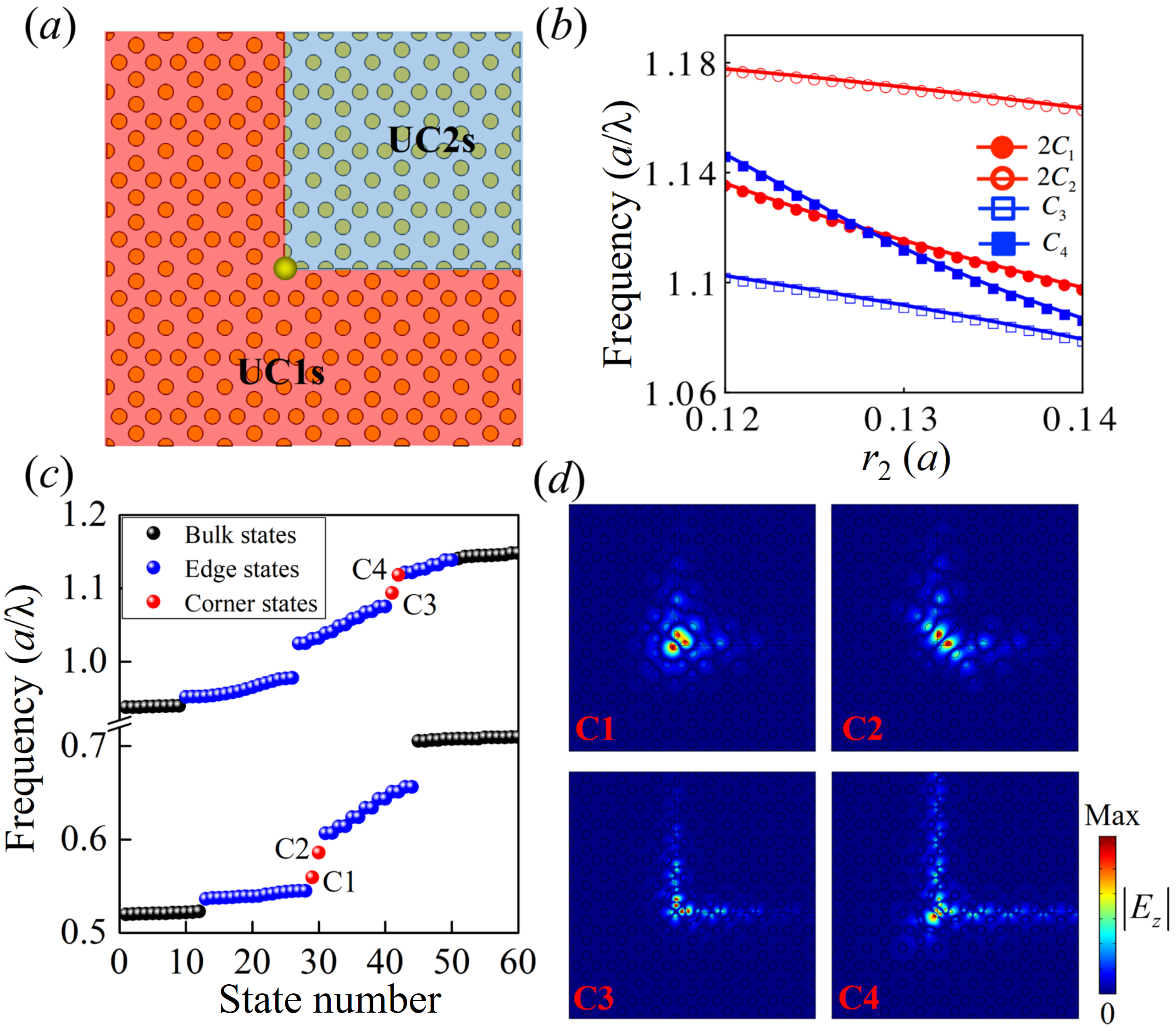} 
\caption{\textit{Emergence of two corner states in both band gaps for SHG.}   (a) 
Schematic of a metastructure supporting corner states, which consists of 
$4\times 4$ UC2s surrounded by three-layer UC1s at the left and bottom sides. 
The yellow ball indicates the location of the corner states. (b) The frequencies of the corner states within the second band gap as a function of $r_2$ with $r_1=0.13a$ and $l=0.29a$. Here
$C_1$ and $C_2$ ($C_3$ and $C_4$) are the two corner states emerging within the first (second) band gap and the frequencies of $C_1$ and $C_4$ can be matched at $r_2=0.128$ for efficient SHG. (c) Calculated eigenfrequencies of the metastructure near the two band gaps at $r_2=0.128$, from which one can see the emergence of two corner states in both band gaps. (d) Eigenmode profiles of the four corner states $C_1$, $C_2$, $C_3$, and $C_4$ in (c). 
 \label{fig:fig6}}
\end{figure}

\section{\label{sec:doubly-corner-state}Emergence of doubly resonant topological corner states for SHG}  
To demonstrate corner states could be induced at a corner between UC1s and UC2s due to the existence of nontrivial topological corner charge, we build a metastructure consisting of $4\times 4$ UC2s surrounded by three-layer of UC1s at the left and bottom sides, as sketched in Fig. \ref{fig:fig6}(a). The yellow ball marks the location of the corner, where nontrivial corner states are expected to emerge. We find that there are two corner states emerging within each band gap (see, e.g., Fig. \ref{fig:fig6}(c)). To demonstrate the frequencies of these corner states could be matched to realize efficient SHG, we show in Fig. \ref{fig:fig6}(b) the evolution of the frequencies of the four corner states within the second band gap as a function of $r_2$ by doubling the frequencies of the two corner states $C_1$ and $C_2$ within the first band gap and presenting them in the second band gap together with $C_3$ and $C_4$. One can see that the frequencies of the corner states of $2C_1$ and $C_4$ can be matched at $r_2=0.128a$. Note here, as the corner states are localized, i.e., their momenta are zero, both the conditions of momentum and frequency matching could be simply realized at $r_2=0.128a$. The calculated eigenfrequencies of the metastructure near the two band gaps at $r_2=0.128a$ are shown in Fig. \ref{fig:fig6}(c), from which one can see the emergence of two corner states within both band gaps (labeled as $C_1$, $C_2$, $C_3$, and $C_4$) apart from the edge and bulk states. The normalized frequencies of these corner states are 0.5596, 0.5862, 1.0942 and 1.1192, respectively, and one can see $2C_1=C_4$ as expected. Figure \ref{fig:fig6}(d) further shows the eigenmode profiles of the four corner states in Fig. \ref{fig:fig6}(c), from which we can see that the corner states are tightly localized around the corner with different eigemode profiles. The Q factors of $C_1$ and $C_4$ are $3.9\times 10^4$ and $5.8\times 10^3$, respectively. One important figure of merit that affects the SHG efficiency is the nonlinear overlap factor between the two interacting modes for FW and SH, which can be defined by \cite{Carletti15OE_overkap},
\begin{gather}
\beta=\frac{\left| \int E^{FF}_zE_z^{FF}\left(E_z^{SH}\right)^*dV\right|}{\sqrt{\int\left|E^{FF}_zE_z^{FF}\right|^2dV}\sqrt{\int\left|E^{SH}_z\right|^2dV}},
\label{OverlapQ}
\end{gather}
where $E_z^{FF}$ and $E_z^{SH}$ represent the electric fields of fundamental wave mode and second harmonic mode, i.e., corner states C1 and C4, respectively, Accordingly, we find that the normalized nonlinear overlap factor between C1 and C4 is 0.123, which is larger than those in \cite{Minkov19Optica_bicSHG,ZinlinSHG}. 
We will show next that the nonlinear interaction of these two corner states with extra-high Q factors for both FW and SH and large spatial overlap factor between them will induce significant enhancement of the SHG efficiency.

 \begin{figure}
\includegraphics[width=\columnwidth]{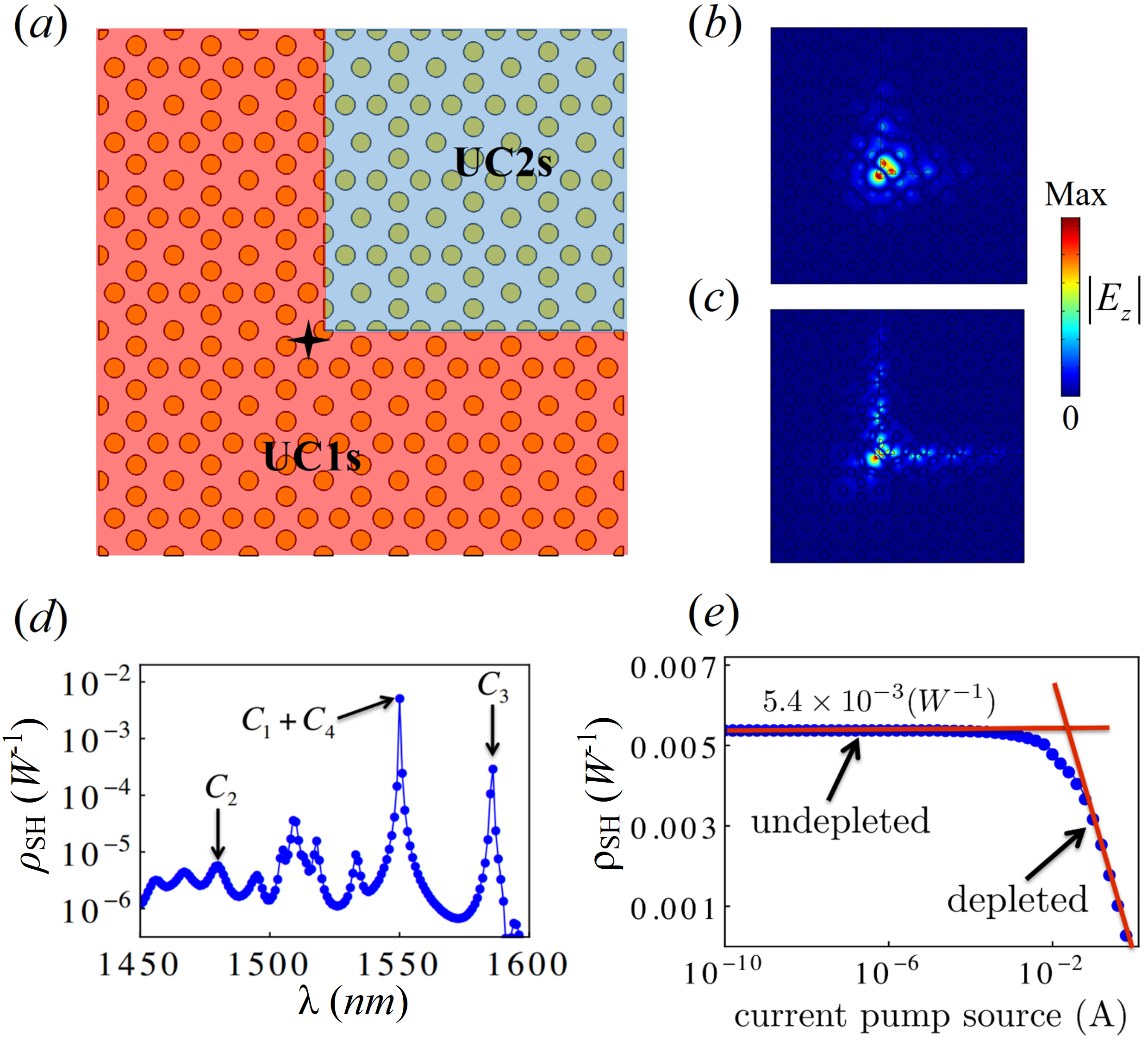} 
\caption{\textit{SHG via doubly resonant corner states.} (a) Schematic of the system for SHG, where the source for excitation of the FW is denoted as the black star. (b) and (c) The electric fields of FW and SH at $\lambda_{\textrm{FW}}=1550nm$, corresponding to the corner states of $C_1$ and $C_4$ in Fig.~\ref{fig:fig2}. (d) The SHG efficiency as a function of the excitation wavelength in the undepleted regime, where the contributions of the four corner states to the enhancement of the efficiency (i.e., peaks) are marked by the arrows and the contributing corner states. (e) The SHG efficiency as a function of pump strength of an out-of-plane line current pump source at $\lambda_{\textrm{FW}}=1550nm$ for the coupling of $C_1$ and $C_4$ corner states in the undepleted and depleted regimes. }
\label{fig:fig7}
\end{figure}

\section{\label{sec:SHG}SHG via doubly resonant corner states}  
To study the SHG due to the nonlinear coupling of the corner states, we launch the FW, i.e., corner state $C_1$ by a source around the corner, as sketched in Fig. \ref{fig:fig7}(a).  Here, for demonstration purpose, the lattice constant is set to $a=867nm$, at which the wavelength of the corner state $C_1$ for FW is $\lambda_{\textrm{FW}} =1550$nm.  Figures \ref{fig:fig7}(b) and (c) show the electric field distributions of the FW and SH with an excitation source of $\lambda_{\textrm{FW}} =1550$nm, respectively, both of which are highly localized around the corner. The fact that both corner states for FW and SH are simultaneously excited can be confirmed by their exited field distributions compared with the eigenmode profiles in Fig. \ref{fig:fig6} (d). Importantly, while the FW is excited by the source, note that there is no source for the SH and the excitation of the SH is purely due to the nonlinear coupling between the two corner states of $C_1$ and $C_4$.

To demonstrate that the  SHG efficiency could be enhanced significantly due to the coupling of the two corner states, we use the intrinsic efficiency of the SHG by \cite{Carletti18PRL_bic} 
\begin{gather}
\rho_{SH}=P_{SH}/P^2_{FW}, 
\end{gather}
where $P_{FW}$ and $P_{SH}$ are the power of the fundamental and harmonic wave, respectively.  The conversion efficiency as a function of the excitation wavelength in the undepleted regime is shown in Fig. \ref{fig:fig7}(d), from which we can see that several peaks emerge. First, the contributions of the four corner states to the enchantment of the SHG efficiency could be clearly identified, which are marked by the arrows and contributing corner states in Fig. \ref{fig:fig7}(d). For example, $C_2$ has a wavelength of $\lambda=1480$nm and $C_3$ has a wavelength of $\lambda=793$nm (i.e., corresponding to FW of $\lambda=1586$nm). Most importantly, the peak at $\lambda=1550$nm corresponds to the coupling of $C_1$ and $C_4$, which has a wavelength of 1550nm and 775nm, respectively. The efficiency due to the coupling of $C_1$ and $C_4$ has a value of $5.4\times 10^{-3} W^{-1}$, which is several orders of magnitude larger than that due to a single bound state in the continuum (BIC) resonance \cite{Koshelev20Science_bic}.  The conversion efficiencies at $\lambda=1480$nm due to $C_2$ and  $\lambda=1585$nm due to $C_3$ are much weaker because only one corner state is involved in the nonlinear process, which is also the recipe that current topological nonlinear optics uses \cite{Kruk20NanoLett_cornerSHG}. Thus this order of magnitude enhancement due to coupling between two corner states illustrates the great prospect of our system to harness high performance SHG for various applications. It is also  worthwhile to discuss the seemingly larger enhancement due to $C_3$ than that due to $C_2$ though in both cases only one corner state is involved in the nonlinear process. For the peak at 1480nm due to $C_2$, from Figs. \ref{fig:fig6} (b) and (c), one can see that the SH frequency at this case is located within the bulk states, thus the SHG is low due to the corner and bulk state coupling in this case. However, for the peak at 1586nm, the nonlinear overlap factor for non-localized SH is calculated to be $2.13\times 10^{-4}$ \cite{supple_material}, which is three orders of magnitude smaller than that at $\lambda = 1550 nm$. Nonetheless, due to the finite width of the resonance at $C_1$, the $C_3$ corner state still has a small overlap with the tail of the $C_1$ resonance, thus the larger enhancement of SHG efficiency compared to the $C_2$ peak.

Finally, we show the SHG efficiency for the coupling of $C_1$ and $C_4$ corner states at excitation wavelength of 1550 nm in Fig.~\ref{fig:fig7}(e) when increasing the pump strength of the source. At low pump strength, one can see the perfect linear scaling between $P_{SH}$ and $P^2_{FW}$, demonstrating the undepleted behaviors. However, at high pump strength, the SHG efficiency begins to decrease due to the   
saturation effect, which also agrees with recent experimental observations \cite{Koshelev20Science_bic}. 

 \begin{figure}
\includegraphics[width=\columnwidth]{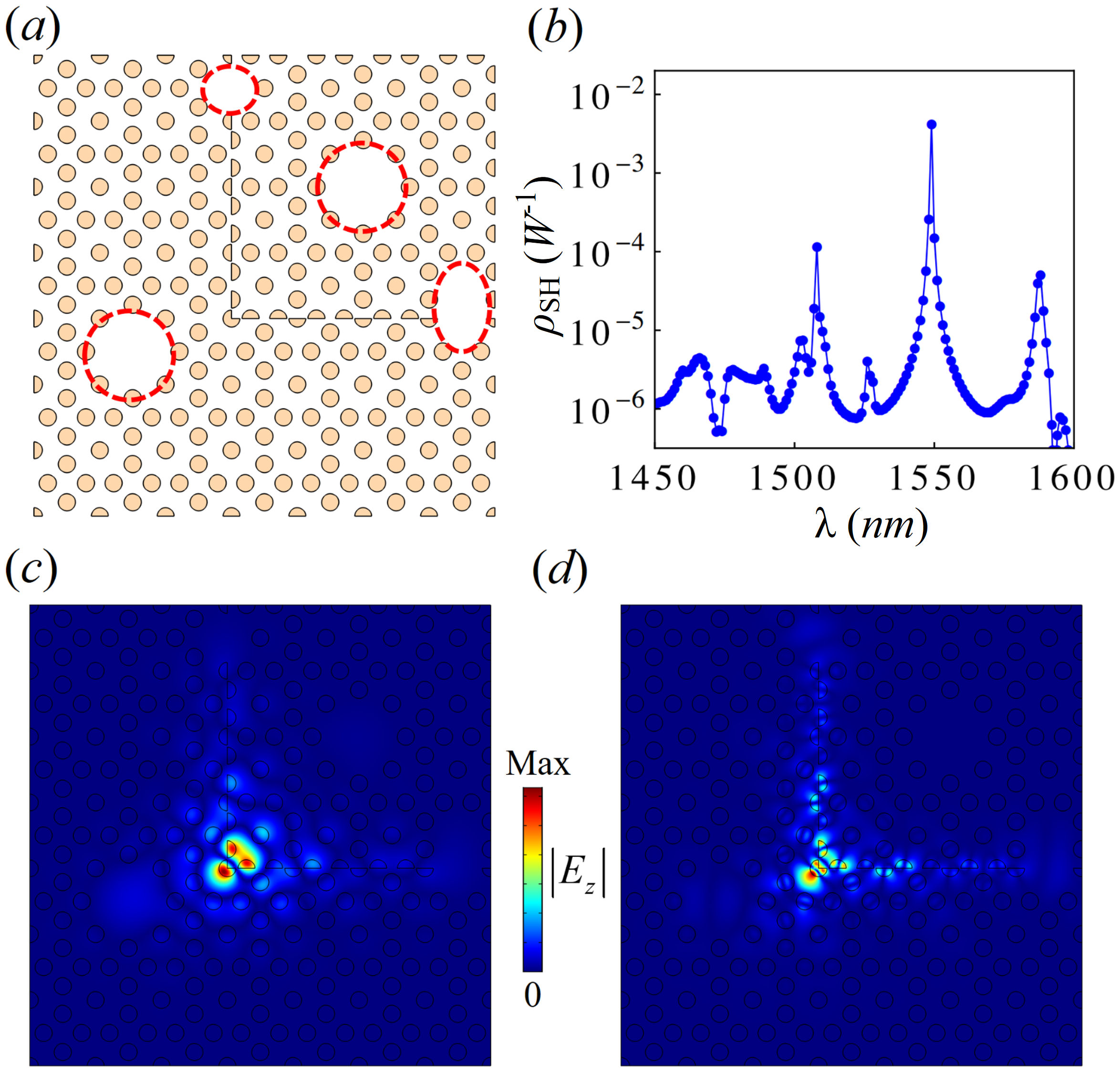} 
\caption{\textit{Robustness of the SHG mediated by doubly resonant corner states.}   (a) Schematic of the system with introduction of defects, where the dielectric materials within the dashed box are removed. (b) The SHG efficiency as a function of the excitation wavelength for this disordered system. (c-d) The electric field distributions of the FW and SH at excitation wavelength of $1550$nm, respectively, corresponding to the corner states of $C_1$ and $C_4$. 
\label{fig:fig8}}
\end{figure}

\section{\label{sec:Robustness}Robustness of the SHG mediated by doubly resonant corner states}  
One of the advantages of topologically protected nonlinear frequency conversion is the robustness of the process against defects. To demonstrate this, we study the SHG of our system after some defects are introduced as shown in Fig. \ref{fig:fig8}(a), where the dielectric materials within the red dashed box are removed. The SHG efficiency as a function of excitation wavelength is presented in Fig. \ref{fig:fig8}(b), from which one can see that the peak at $1550$nm due to the coupling of corner states $C_1$ and $C_4$ is not changed whereas other parts of the curve are increased or decreased due to randomness of the defects introduced. We further show in  Figs. \ref{fig:fig8}(c) and (d) the electric field distributions of the FW and SH at excitation wavelength of $1550$nm, corresponding to the coupling of the two corner states $C_1$ and $C_4$. Compared with the electric field distributions in Figs. \ref{fig:fig7}(b) and (c) of the unperturbed system or the eigenmode profiles in Fig. \ref{fig:fig6}(d), one can see that the coupling of the two corner states is not destroyed due to the introduction of defects, thus demonstrating that our system on SHG mediated by doubly resonant corner states is indeed robust against defects and this justifies the topologically protected nature of this kind of nonlinear process. In Supplementary Material~\cite{supple_material}, we further demonstrate that the system is robust against different defects. For experimental implementations, the setup we proposed could be fabricated following the recent experimental works \cite{Smirnova19PRL_THG, Kruk20NanoLett_cornerSHG}, where the fundamental corner mode could be excited by a tunable continuous-wave laser source, collimated and focused on the corner by a microscope objective. The SHG signal could be collected through the same objective, redirected by a dichroic mirror to separate the pump beam from the harmonic signal.  Furthermore, the electric fields at both fundamental and harmonic frequencies could be greatly enhanced due to the field confinement effects, which could in turn reduce the power needed for the experimental observations.

\section{\label{sec:Conclusion}Conclusion and outlook}  
In conclusion, we have demonstrated giant enhancement of SHG efficiency using doubly resonant photonic topological corner modes, which bridges nonlinear topological photonics with high-order topological photonics and is expected to generate broad impact in both areas. The concept presented here could also be extended to other nonlinear optical processes, such as third harmonic generation or four-wave mixing, e.g., it would be interesting to see whether one could match the frequencies of the four corner states in the two band gaps for enhancement of  four-wave mixing related applications. Moreover, it is certainly interesting to see whether nonlinear frequency conversion could be achieved via nonlinear interaction of other high-order photonic topological modes, such as third-order corner states or high-order hinge states in 3D~\cite{Kim20Light} or even higher-order topological corner states in the continuum~\cite{Cerjan20PRL_BIC}.

\textit{Acknowledgments.---} This work is supported by National Natural Science Foundation of China (No.1210020421), the Hong Kong Scholars Program (No. XJ2020004) and the Research Grants Council of Hong Kong SAR (Grant No. C6013-18G and AoE/P-502/20).


\begin{thebibliography}{99}

\bibitem{Lu14NP_review}  L. Lu, J.D. Joannopoulos, M. Soljacic, {\it Topological photonics}, Nat. Photonics {\bf 8}, 821 (2014).
\bibitem{Lu16NP}  L. Lu, J. D. Joannopoulos, and M. Soljacic, {\it Topological states in photonic systems}, Nat. Phys. {\bf 12}, 626 (2016).
\bibitem{Khanikaev17NatPho_2Dtopo} A. B. Khanikaev and G. Shvets, {\it Two-dimensional topological photonics}, Nat. Photonics {\bf 11}, 763 (2017).
\bibitem{Ozawa19RMP} T. Ozawa, H. M. Price, A. Amo, N. Goldman, M. Hafezi, L. Lu, M. C. Rechtsman, D. Schuster, J. Simon, and O. Zilberberg, {\it Topological photonics},  Rev. Mod. Phys. {\bf 91}, 015006 (2019).
\bibitem{Khanikaev13NM_PTI}  A. B. Khanikaev, S.H. Mousavi, W.-K. Tse, M. Kargarian, A. H. MacDonald, and G. Shvets, {\it Photonic topological insulators}, Nat. Mater. {\bf 12}, 233 (2013).
\bibitem{WuHu15PRL} L.-H. Wu and X. Hu, {\it Scheme for achieving a topological photonic crystal by using dielectric material}, Phys. Rev. Lett. {\bf 114}, 223901 (2015) .
\bibitem{Hang18PRL_ExpSpin} Y. Yang, Y. F. Xu, T. Xu, H.-X. Wang, J.-H. Jiang, X. Hu, and Z. H. Hang, {\it Visualization of a unidirectional electromagnetic waveguide using topological photonic crystals made of dielectric materials}, Phys. Rev. Lett. {\bf 120}, 217401 (2018).
\bibitem{Bahari17Science_lasing} B. Bahari, A. Ndao, F. Vallini, A. El Amili, Y. Fainman, and B. Kante, {\it Nonreciprocal lasing in topological cavities of arbitrary geometries}, Science {\bf 358}, 636 (2017).
\bibitem{yafeng RRL}  Y. Chen, F. Meng, G.Li, B. Jia, and X. Huang, {\it Inverse Design of Photonic Topological Insulators with Extra‐Wide Bandgaps},Phys Status Solidi-R. {\bf 13}, 1900175 (2019).
\bibitem{Harari18Science_TLlaserT} G. Harari, M. A. Bandres, Y. Lumer, M. C. Rechtsman, Y. D. Chong, M. Khajavikhan, D. N. Christodoulides, and M. Segev, {\it Topological insulator laser: theory}, Science {\bf 359}, eaar4003 (2018).
\bibitem{Zeng20Nature_valleylaser}  Y. Zeng, U. Chattopadhyay, B. Zhu, B. Qiang, J. Li, Y. Jin, L. Li, A. G. Davies, E. H. Linfield, B. Zhang, Y. Chong, and Q. J. Wang, {\it Electrically pumped topological laser with valley edge modes}, Nature {\bf 578}, 246 (2020).

\bibitem{Xie18PRB_corner}  B.-Y. Xie, H.-F. Wang, H.-X. Wang, X.-Y. Zhu, J.-H. Jiang, M.-H. Lu, and Y.-F. Chen, {\it Second-order photonic topological insulator with corner states}, Phys. Rev. B {\bf 98}, 205147 (2018).
\bibitem{Xie19PRL_exp}  B.-Y. Xie, G.-X. Su, H.-F. Wang, H. Su, X.-P. Shen, P. Zhan, M.-H. Lu, Z.-L. Wang, and Y.-F. Chen, {\it Visualization of higher-order topological insulating phases in two-dimensional dielectric photonic crystals}, Phys. Rev. Lett. {\bf 122},  233903 (2019).
\bibitem{Chen19PRL_exp}  X.-D. Chen, W.-M. Deng, F.-L. Shi, F.-L. Zhao, M. Chen, and J.-W. Dong, {\it Direct observation of corner states in second-order topological photonic crystal slabs}, Phys. Rev. Lett. {\bf 122}, 233902 (2019).
\bibitem{Hassan19NatPho_coner}  A. El Hassan, F. K. Kunst, A. Moritz, G. Andler, E. J. Bergholtz, and M. Bourennane, {\it Corner states of light in photonic waveguides}, Nat. Photonics {\bf 13}, 697 (2019).
\bibitem{Chen20PRR_inverse}  Y. Chen, F. Meng, Y. Kivshar, B. Jia, and X. Huang, {\it Inverse design of higher-order photonic topological insulators}, Phys. Rev. Research {\bf 2}, 023115 (2020).
\bibitem{Kim20Light}  M. Kim, Z. Jacob, J. Rho, {\it Recent advances in 2D, 3D and higher-order topological photonics}, Light Sci. Appl. {\bf 9}, 130 (2020).
\bibitem{Mittal10NatPho_quadrupole}  S. Mittal, V. V. Orre, G. Zhu, M. A. Gorlach, A. Poddubny, and M. Hafezi, {\it Photonic quadrupole topological phases}, Nat. Photonics {\bf 13}, 692 (2019).
\bibitem{He20NatCom_Quadrupole}  L. He, Z. Addison, E.J. Mele, and B. Zhen, {\it Quadrupole topological photonic crystals}, Nat. Commun. {\bf 11},  3119 (2020).
\bibitem{Zhang20AdvSci}  L. Zhang, Y. Yang, Z.-K. Lin, P. Qin, Q. Chen, F. Gao, E. Li, J.-H. Jiang, B. Zhang, and H. Chen, {\it Higher-Order Topological States in Surface-Wave Photonic Crystals}, Adv. Sci. {\bf 7},  1902724 (2020).
\bibitem{Ota19Optica_cornercavity}  Y. Ota, F. Liu, R. Katsumi, K. Watanabe, K. Wakabayashi, Y. Arakawa, and S. Iwamoto, {\it Photonic crystal nanocavity based on a topological corner state}, Optica {\bf 6}, 786 (2019).
%
\bibitem{Zhang20Light_cornerlaser}  W. Zhang, X. Xie, H. Hao, J. Dang, S. Xiao, S. Shi, H. Ni, Z. Niu, C. Wang, K. Jin, X. Zhang, and X. Xu, {\it Low-threshold topological nanolasers based on the second-order corner state}, Light Sci. Appl. {\bf 9}, 109 (2020).
\bibitem{Kim20NatCom_conerlasing}  H.-R. Kim, M.-S. Hwang, D. Smirnova, K.-Y. Jeong, Y. Kivshar, and H.-G. Park, {\it Multipolar lasing modes from topological corner states}, Nat. Commun. {\bf 11}, 5758 (2020).

\bibitem{Smirnova20APR_NTPreview}  D. Smirnova, D. Leykam, Y. Chong, and Y. Kivshar, {\it Nonlinear topological photonics}, Appl. Phys. Rev. {\bf 7}, 021306 (2020).
\bibitem{Dobrykh18PRL_nonlinear}  D. A. Dobrykh, A.V. Yulin, A. P. Slobozhanyuk, A. N. Poddubny, and Y. S. Kivshar, {\it Nonlinear Control of Electromagnetic Topological Edge States}, Phys. Rev. Lett.  {\bf 121}, 163901 (2018).
\bibitem{Smirnova19PRL_THG}  D. Smirnova, S. Kruk, D. Leykam, E. Melik-Gaykazyan, D.-Y. Choi, and Y. Kivshar, {\it Third-Harmonic Generation in Photonic Topological Metasurfaces}, Phys. Rev. Lett.  {\bf 123}, 103901 (2019).

\bibitem{Mukherjee_Science20}  S. Mukherjee and M. C. Rechtsman, {\it Observation of Floquet solitons in a topological bandgap}, Science {\bf 368}, 856 (2020).
\bibitem{Maczewsky_Science20}  L. J. Maczewsky, M. Heinrich, M. Kremer, S. K. Ivanov, M. Ehrhardt, F. Martinez, Y. V. Kartashov, V. V. Konotop, L. Torner, D. Bauer, and A. Szameit, {\it Nonlinearity-induced photonic topological insulator}, Science {\bf 370}, 701 (2020).
\bibitem{Xia_Science21}  S. Xia, D. Kaltsas, D. Song, I. Komis, J. Xu, A. Szameit, H. Buljan, K. G. Makris, and Z. Chen {\it Nonlinear tuning of PT symmetry and non-Hermitian topological states}, Science {\bf 372}, 72 (2021).

\bibitem{Jurgensen_arxiv21_Thouless}  M. Jurgensen, S. Mukherjee, and M. Rechtsman, {\it Quantized Nonlinear Thouless Pumping}, ArXiv 2106.14128 (2021).

\bibitem{Boyd08Book}  R. Boyd, {\it Nonlinear Optics (Third Edition)}, Academic Press, Burlington, 2008.
\bibitem{Kruk19NatNaT_nonlinear}  S. Kruk, A. Poddubny, D. Smirnova, L. Wang, A. Slobozhanyuk, A. Shorokhov, I. Kravchenko, B. Luther-Davies, and Y. Kivshar, {\it Nonlinear light generation in topological nanostructures}, Nat. Nanotechnol. {\bf 14}, 126 (2019). 
\bibitem{Wang19NatCom_harmonic}  Y. Wang, L.-J. Lang, C.H. Lee, B. Zhang, and Y. D. Chong, {\it Topologically enhanced harmonic generation in a nonlinear transmission line metamaterial}, Nat. Commun. {\bf 10}, 1102 (2019).
\bibitem{Qian18OE_doubleedge}  C. Qian, K. H. Choi, R. P. H. Wu, Y. Zhang, K. Guo, and K. H. Fung, {\it Nonlinear frequency up-conversion via double topological edge modes}, Opt. Express {\bf 26},  5083 (2018).
%
\bibitem{You20SciAdv}  J. W. You, Z. Lan, and N.C. Panoiu, {\it Four-wave mixing of topological edge plasmons in graphene metasurfaces}, Sci. Adv. {\bf 6}, eaaz3910 (2020).
\bibitem{Lan21PRA_valleySHG}  Z. Lan, J. W. You, Q. Ren, W. E. I. Sha, and N. C. Panoiu, {\it Second-harmonic generation via double topological valley-Hall kink modes in all-dielectric photonic crystals}, Phys. Rev. A {\bf 103}, L041502 (2021).
\bibitem{Lan20PRB_nonlinear}  Z. Lan, J.W. You, and N.C. Panoiu, {\it Nonlinear one-way edge-mode interactions for frequency mixing in topological photonic crystals}, Phys. Rev. B {\bf 101}, 155422 (2020).
\bibitem{Carletti18PRL_bic}  L. Carletti, K. Koshelev, C. De Angelis, and Y. Kivshar, {\it Giant nonlinear response at the nanoscale driven by bound states in the continuum}, Phys. Rev. Lett. {\bf 121}, 033903 (2018).
\bibitem{Koshelev20Science_bic}  K. Koshelev, S. Kruk, E. Melik-Gaykazyan, J.-H. Choi, A. Bogdanov, H.-G. Park, and Y. Kivshar, {\it Subwavelength dielectric resonators for nonlinear nanophotonics}, Science {\bf 367}, 288 (2020).
\bibitem{Wang20Optica_bicSHG}  J. Wang, M. Clementi, M. Minkov, A. Barone, J.-F. Carlin, N. Grandjean, D. Gerace, S. Fan, M. Galli, and R. Houdre, {\it Doubly resonant second-harmonic generation of a vortex beam from a bound state in the continuum}, Optica {\bf 7},1126  (2020).
\bibitem{Minkov19Optica_bicSHG}  M. Minkov, D. Gerace, S. Fan, {\it Doubly resonant nonlinear photonic crystal cavity based on a bound state in the continuum}, Optica {\bf 6}, 1039 (2019).
\bibitem{ZinlinSHG}  Z. Lin, X. Liang, M. Lončar, SG. Johnson, AW Rodriguez, {\it Cavity-enhanced second-harmonic generation via nonlinear-overlap optimization}, Optica {\bf 3}, 233 (2016).
\bibitem{Kruk20NanoLett_cornerSHG}  S. S. Kruk, W. Gao, D.-Y. Choi, T. Zentgraf, S. Zhang, and Y. Kivshar, {\it Nonlinear Imaging of Nanoscale Topological Corner States}, Nano Lett. {\bf 21}, 4592 (2021).

\bibitem{Liu17PRL_zeroBC}  F. Liu, K. Wakabayashi, {\it Novel Topological Phase with a Zero Berry Curvature}, Phys. Rev. Lett. {\bf 118}, 076803 (2017).
\bibitem{Chen19PRB_cornerbic}  Z.-G. Chen, C. Xu, R. Al Jahdali, J. Mei, and Y. Wu, {\it Corner states in a second-order acoustic topological insulator as bound states in the continuum}, Phys. Rev. B {\bf 100}, 075120 (2019).
\bibitem{Meng20APL_3D}  Y. Chen,F. Meng, Z. Lan, B. Jia, and X. Huang, {\it Dual-Polarization Second-Order Photonic Topological Insulators},Phys. Rev. Appl. {\bf 15}, 034053 (2021).
\bibitem{Liu19PRL_Helical}  F. Liu, H.-Y. Deng, K. Wakabayashi, {\it Helical Topological Edge States in a Quadrupole Phase}, Phys. Rev. Lett. {\bf 122}, 086804 (2019).
\bibitem{Zhang19PRL_nonHsonic}  Z. Zhang, M. R. Lopez, Y.,  Cheng, X. Liu, and J. Christensen, {\it Non-Hermitian sonic second-order topological insulator}, Phys. Rev. Lett. {\bf 122}, 195501 (2019).
\bibitem{Zhang19AdvMat_Acoustic}  Z. Zhang, H. Long, C. Liu, C. Shao, Y. Cheng, X. Liu, and J. Christensen, {\it Deep-Subwavelength Holey Acoustic Second-Order Topological Insulators}, Adv. Mater. {\bf 31}, 1904682 (2019).
\bibitem{Carletti15OE_overkap}L. Carletti, A. Locatelli, O. Stepanenko, G. Leo, C. De Angelis, {\it Enhanced second-harmonic generation from magnetic resonance in AlGaAs nanoantennas}, Opt. Express {\bf 23} 26544, (2015).\bibitem{supple_material} See Supplemental Material at [XXX] for details of topological edge states and further discussions on robustness.
\bibitem{Cerjan20PRL_BIC}A. Cerjan, M. Jurgensen, W. A. Benalcazar, S. Mukherjee, and M. C. Rechtsman, {\it Observation of a Higher-Order Topological Bound State in the Continuum}, Phys. Rev. Lett. {\bf 125}, 213901 (2020). 



\end{thebibliography}
\end{document}